\documentclass[oldversion]{aa}
\usepackage{graphicx}
\usepackage{natbib}
\newcommand{\degree}{^{\circ}}

\begin{document}

   \title{The AGILE observations of the hard and bright GRB~100724B}

   \author{E.~Del~Monte\inst{1},
   G.~Barbiellini\inst{2,3}, I.~Donnarumma\inst{1},
   F.~Fuschino\inst{4},  A.~Giuliani\inst{5}, F.~Longo\inst{2,3}, M.~Marisaldi\inst{4}, G.~Pucella\inst{6},
   M.~Tavani\inst{1,7}, M.~Trifoglio\inst{4}, A.~Trois\inst{1},
   A.~Argan\inst{1}, A.~Bulgarelli\inst{4},
   P.~Caraveo\inst{5}, P.W.~Cattaneo\inst{8}, A.W.~Chen\inst{5},
   E.~Costa\inst{1}, F.~D'Ammando\inst{9}, G.~Di~Cocco\inst{4},
   Y.~Evangelista\inst{1}, M.~Feroci\inst{1},
   M.~Galli\inst{10}, F.~Gianotti\inst{4},  C.~Labanti\inst{4},
   I.~Lapshov\inst{1}, F.~Lazzarotto\inst{1}, P.~Lipari\inst{11},
   S.~Mereghetti\inst{5}, E.~Moretti\inst{2,3}, A.~Morselli\inst{7},
   L.~Pacciani\inst{1}, A.~Pellizzoni\inst{12}, F.~Perotti\inst{5},
   G.~Piano\inst{1}, P.~Picozza\inst{7}, M.~Pilia\inst{12,13},
   M.~Prest\inst{13}, M.~Rapisarda\inst{6},
   A.~Rappoldi\inst{8}, S.~Sabatini\inst{1}, P.~Soffitta\inst{1},
   E.~Striani\inst{1}, E.~Vallazza\inst{2}, S.~Vercellone\inst{9}, V.~Vittorini\inst{1},
   L.~A.~Antonelli\inst{14,15}, S.~Cutini\inst{14,16}, C.~Pittori\inst{14,16}, P.~Santolamazza\inst{14,16},
   F.~Verrecchia\inst{14,16}, P.~Giommi\inst{14,17}, L.~Salotti\inst{17}}

   \offprints{E. Del Monte}

   \institute{INAF IASF Roma, Via Fosso del Cavaliere 100, I-00133 Roma, Italy\\ 
              \email{ettore.delmonte@iasf-roma.inaf.it}
              \and
              INFN Trieste, Padriciano 99, I-34012 Trieste, Italy 
              \and
              Dip. di Fisica, Universit\`a di Trieste, Via Valerio 2, I-34127 Trieste, Italy
              \and
              INAF IASF Bologna, Via Gobetti 101, I-40129 Bologna, Italy 
              \and
              INAF IASF Milano, Via E. Bassini 15, I-20133 Milano, Italy 
              \and
              ENEA C. R. Frascati, Via E. Fermi 45, I-00044 Frascati (Rm), Italy 
              \and
              Dip. di Fisica, Universit\`a degli Studi di Roma ``Tor Vergata'',  Via della Ricerca Scientifica 1, I-00133 Roma, Italy 
              \and
              INFN Pavia, Via Bassi, 6 I-27100 Pavia, Italy 
              \and
              INAF IASF Palermo, Via U.\ La Malfa 153, I-90146 Palermo, Italy 
              \and
              ENEA C.R. ``E. Clementel'', Via Martiri di Monte Sole 4, I-40129 Bologna, Italy 
              \and
              Dip. di Fisica, Universit\`a degli Studi di Roma ``La Sapienza'', P.le A. Moro 5, I-00185 Roma, Italy 
              \and
              INAF Osservatorio Astronomico di Cagliari, loc. Poggio dei Pini, strada 54, I-09012, Capoterra (Ca), Italy 
              \and
              Dip. di Fisica e Matematica, Universit\`a dell'Insubria, Via Valleggio 11, I-20100 Como, Italy 
              \and
              ASI Science Data Center, Via G.\ Galilei, I-00044 Frascati (Rm), Italy 
              \and
              INAF Osservatorio Astronomico di Roma, Via di Frascati 33, I-00040 Monte Porzio Catone (Rm), Italy 
              \and
              INAF staff resident at ASI Science Data Center 
              \and
              Agenzia Spaziale Italiana, Unit\`a Osservazione dell'Universo, Viale Liegi 26, 00198 Roma, Italy 
              }

   \date{}

\abstract{The observation of gamma ray bursts (GRBs) in the gamma
ray band has been advanced by the \textit{AGILE} and
\textit{Fermi} satellites after the era of the \textit{Compton
Gamma-Ray Observatory}. \textit{AGILE} and \textit{Fermi} are
showing that the GeV-bright GRBs share a set of common features,
particularly the high fluence from the keV up to the GeV energy
bands, the high value of the minimum Lorentz factor, an extended
emission of gamma rays, which is often delayed with respect to
lower energies, and finally the possible multiple spectral
components. GRB 100724B, localised in a joint effort by
\textit{Fermi} and the InterPlanetary Newtork, is the brightest
burst detected in gamma rays so far by \textit{AGILE}.
Characteristic features of GRB 100724B are the simultaneous
emissions at MeV and GeV, without delayed onset or any time lag as
shown by the analysis of the cross correlation function, and the
significant spectral evolution in hard X-rays over the event
duration. In this paper we show the analysis of the \textit{AGILE}
data of GRB 100724B and discuss its features in the context of the
bursts observed so far in gamma rays and the recently proposed
models.}


\authorrunning{E. Del Monte et al.}
\titlerunning{GRB~100724B observed by AGILE}

\maketitle

\section{Introduction}

In more than thirty years of study a wealth of information has
been gathered about the prompt emission of gamma ray bursts (GRBs)
in hard X-rays. The interested reader may find recent and
comprehensive reviews by \citet{Piran_2004},
\citet{Meszaros_2006}, and \citet{Gehrels_et_al_2009}, among
others. In comparison, before the launch of the \textit{AGILE}
\citep{Tavani_et_al_2009} and \textit{Fermi}
\citep{Michelson_Atwood_Ritz_2010} satellites, limited information
was available about the GRB higher energy component. In fact, the
EGRET spark chamber \citep{Kanbach_et_al_1988} aboard the
satellite-borne \textit{Compton Gamma-Ray Observatory} only
detected a small number of GRBs in the energy band between
hundreds of MeV and few GeV: GRB 910503
\citep{Schneid_et_al_1992}, GRB 910601 \citep{Dingus_1995}, GRB
910814 \citep{Schaefer_et_al_1998}, GRB 920622
\citep{Schneid_et_al_1995}, GRB 930131
\citep{Kouveliotou_et_al_1994,Sommer_et_al_1994}, GRB 940217
\citep{Hurley_et_al_1994}, and GRB 940301
\citep{Schneid_et_al_1995,Dingus_1995}. Combined spectra of a
sample of 15 GRBs detected by the \textit{Compton Gamma-Ray
Observatory} are analysed by \citet{Kaneko_et_al_2008} in an
energy range between tens of keV \citep[from
BATSE,][]{Fishman_et_al_1992} and tens of MeV (from the EGRET TASC
calorimeter). EGRET also discovered that, compared with the keV
and MeV energy bands, the gamma ray emission may last much longer,
with the exceptional case of GRB 940217 \citep{Hurley_et_al_1994},
whose gamma ray component lasted for $\simeq 5400$ s and from
which a photon of 18 GeV was detected $\simeq 5000$ s after the
trigger.

In recent years, after the demise of the \textit{Compton Gamma-Ray
Observatory}, the observation of GRBs in the MeV to GeV band is
continuing thanks to the gamma ray imagers aboard the
\textit{AGILE} and \textit{Fermi} satellites. GRBs observed in
gamma rays by \textit{AGILE} and \textit{Fermi}, which include
both long and short events, exhibit peculiar and distinctive
features. The gamma ray emission generally starts during the
prompt phase and may be simultaneous with the hard X-ray
component, as in GRB 090217A \citep{Ackermann_et_al_090217A}, or
may show a delay, with the most striking case being the short GRB
090510
\citep{Giuliani_et_al_2010,Ackermann_et_al_090510,Kumar_Barniol-Duran_2010}.
Moreover, in the majority of GRBs observed so far in gamma rays,
the GeV component has a longer duration than the keV -- MeV one. A
recent and detailed review of the GRB features in gamma rays is
provided, for example, by \citet{Zhang_2011}.

The \textit{Fermi} observation of GRB 090902B
\citep{Abdo_et_al_090902B} is a paradigmatic example of detecting
different components in the spectrum of a gamma ray bright GRB. We
recall here that the GRB spectral energy distribution can be
modelled using the phenomenological Band function
\citep[][]{Band_et_al_1993}, and on average the parameters show a
peak energy around $\simeq 300$ keV and photon indices $\simeq
-1.1$ and $\simeq -2.3$ for the low and high energy powerlaws,
respectively \citep{Preece_et_al_2000}. The spectrum of GRB
090902B can be modelled using two components: the Band function in
the energy band between 50 keV and 40 MeV and an additional
powerlaw, fitting simultaneously the emission below 50 keV and
above 100 MeV. In other events, such as GRB 080514B observed by
\textit{AGILE} \citep{Giuliani_et_al_2008} and GRB 080916C studied
by \textit{Fermi} \citep{Abdo_et_al_080916C}, the spectral shape
is the same from keV up to GeV energies, and a single Band
function is an adequate model for the whole spectrum.

Some hints that the high fluence is an inherent peculiarity of the
GeV-emitting GRBs date back to the EGRET era. For example, GRB
940217 has a remarkable fluence of $(6.6 \pm 2.7) \times 10^{-4}
\; \mathrm{erg \; cm^{-2}}$ above 20 keV
\citep{Hurley_et_al_1994}. From the analysis of a sample of twelve
GRBs detected in the GeV band by the \textit{Fermi} Large Area
Telescope, \citet{Ghisellini_et_al_2010} find that a common
property is their high fluence measured in the 8 keV -- 10 MeV
energy band by the Gamma-ray Burst Monitor, belonging to the $2
\sigma$ (95~\%) to $3 \sigma$ (99.7 \%) highest tail in the
distribution from the catalogue of 121 GRBs of the latter
instrument (until October 2009).

Another important characteristic of GRBs detected in gamma rays is
the remarkably high value of their bulk Lorentz factor. Since a
spectral cutoff is not detected in the GeV interval, the optical
depth of the interactions of gamma rays on low-energy photons is
small \citep{Baring_Harding_1997,Lithwick_Sari_2001}. The minimum
bulk Lorentz factor can consequently be estimated based on the
maximum detected energy and the time variability, on the
hypothesis that the photon number is distributed following a
powerlaw, as reported on the supporting online material of the
paper by \citet{Abdo_et_al_080916C}. Using this method, minimum
Lorentz factor values ranging from hundreds up to about one
thousand \citep[for GRB 090510,][]{Ackermann_et_al_090510} are
derived for gamma ray bright GRBs. For comparison, using a similar
method \citep[discussed by][]{Lithwick_Sari_2001},
\citet{Rossi_et_al_2010} estimate a minimum of a few hundreds on
the initial bulk Lorentz factor of the non-GeV emitting GRB
080928, detected up to 150 keV by \textit{Fermi} and
\textit{Swift} \citep{Gehrels_et_al_2004}. However, the estimation
of the minimum Lorentz factor depends on the model of the photon
density in the burst emission mechanism and values lower by a
factor of two to three are found, for example, by
\citet{Hascoet_et_al_2011}.

The first detections of GeV photons from GRBs by \textit{AGILE}
and \textit{Fermi} have raised a debate about the emission
mechanisms. A complete and exhaustive review of all the available
models is far beyond the scope of this paper. In the following, we
concentrate mainly on the dichotomy between internal shock (i. e.
prompt emission) and external shock (i. e. afterglow emission)
scenarios to explain the peculiar characteristics of the GRB gamma
ray component, especially the delayed onset and the longer
duration with respect to lower energies, the overall spectral
shape and the possible presence of additional spectral components.
\citet{Ghisellini_et_al_2010} discuss the properties of twelve
GRBs detected by \textit{Fermi} and interpret the gamma ray
emission as afterglow, particularly on the basis of the temporal
decay slope, similar to that of afterglows observed in X-rays.
Following their interpretation, the decay slope is also consistent
with a radiative regime, in which all the dissipated energy is
radiated away. The gamma ray emission is also interpreted as
afterglow by \citet{Kumar_Barniol-Duran_2010}, who establish a
correlation between the gamma ray flux (above 100 MeV) and the
flux of the X-ray and optical afterglows at a late time. From the
synchrotron emission model, the authors can estimate the
properties of the external shock emission, such as the fraction of
energy in electrons and magnetic field, the circumstellar medium
density, and the burst kinetic energy, consistently in both the
prompt flux above 100 MeV and the late-time X-ray and optical
afterglows. The external shock scenario is challenged, for
example, by \citet{Maxham_Zhang_Zhang_2011} based on the fact
that, above 100 MeV, the external shock model that fits the data
at a late time (50 -- 500 s after trigger) gives just a fraction
of the flux at an early time (1 -- 50 s after trigger). The
authors consequently interpret the gamma ray flux as mainly due to
the internal shock, with only a small contribution from the
external shock.

The paper is organised as follows: the detection of GRB 100724B is
described in section \ref{sec:Discovery}, the analysis of the
\textit{AGILE} data is reported in section
\ref{sec:AGILE_observation}, and finally we discuss our results
and draw our conclusions in section \ref{sec:Discussion}.

\section{Detection of GRB 100724B}
\label{sec:Discovery}

GRB 100724B was localised by the Gamma-ray Burst Monitor
\citep[GBM;][]{Meegan_et_al_2009} aboard \textit{Fermi} on 24 July
2010 00:42:05.98 UT at the position $RA = 124.16 \degree$, $dec =
74.42 \degree$ with a statistical error of $1.0 \degree$ and a
systematic error in the range between $2 \degree$ and $3 \degree$
\citep{GCN_10977}. The \textit{Fermi} Large Area Telescope
\citep[LAT;][]{Atwood_et_al_2009} localised the GRB at a position
($RA = 120.04 \degree$, $dec = 76.74 \degree$) consistent with GBM
and with an error box of $1.1 \degree$ \citep{GCN_10978}. The
position with the smallest error region is a strip in the sky with
$1.2 \degree$ length and $0.2 \degree$ width centred on $RA =
118.8 \degree$, $dec = 75.8 \degree$
\citep[][]{Guiriec_et_al_2011}, which is obtained from the
intersection of the  LAT  error box at 90 \% confidence level and
the annulus produced by the InterPlanetary Network (IPN) with the
GBM, Konus-Wind, and MESSENGER data.

GRB 100724B also triggered \citep{GCN_10994} the \textit{AGILE}
Minicalorimeter \citep[MCAL;][]{Labanti_et_al_2009} on 24 July
2010 00:42:00 UT, hereafter considered as the event trigger time
($t_0$). The burst was also detected \citep{GCN_10996} by the
\textit{AGILE} Gamma Ray Imaging Detector
\citep[GRID;][]{Prest_et_al_2003} with a dedicated analysis aimed
at finding the counterpart of GRBs localised in hard X-rays by
other instruments \citep[see][for a description of the
technique]{Moretti_et_al_2009}. At the trigger time only some of
the \textit{AGILE} contacts with the Malindi ground station were
available for the telemetry downlink, consequently the hard X-ray
imager SuperAGILE \citep[see][for a
description]{Feroci_et_al_2007} was configured not to collect data
in order to ensure the complete telemetry storage by the GRID.

At the time of the GBM detection, the GRB distance from the Sun
was only $54.5 \degree$ and the \textit{Swift} satellite could not
perform follow-up observations owing to Sun aspect
constraints. The GRB position went out from the \textit{Swift} Sun
aspect constraints on 2 August 2010 and two follow-up observations
were performed on slightly different positions, starting $8.41
\times 10^5$ s (with observation id 20144) and $8.47 \times 10^5$
s (with observation id 20145) after trigger respectively. No
afterglow candidates are found in the two pointings and no more
follow-up observation were performed due to the tight schedule of
the \textit{Swift} mission, thus GRB 100724B lacks an associated
afterglow and a redshift.

\section{The AGILE observation of the prompt emission}
\label{sec:AGILE_observation}

\subsection{Data reduction}

The MCAL data are stored in photon-by-photon mode in a temporary
buffer and are provided to telemetry only on trigger
\citep[see][for a description]{Fuschino_et_al_2008}. In the case
of GRB 100724B, the buffer contains data from $\simeq 40$~s before
until $\simeq 200$ s after trigger. We did not detect any GRB
precursor, at a significance level of $5\sigma$, in the available
MCAL data before trigger, thus all these events are used to
estimate the instrument background. We selected the MCAL data
above 0.3 MeV energy, thus ensuring a negligible contamination
from the instrumental electronic noise.

Since January 2010, \textit{AGILE} is operating in a spinning mode
with an angular velocity of $\simeq 0.8 \degree$ per second around
the axis pointed toward the Sun. The spinning mode produces the
apparent motion of the sources in the GRID field of view (FoV) and
the ``transit'' duration depends on the source position and the
satellite attitude. In the case of GRB 100724B, the distance
between the source position and the satellite boresight is $\simeq
65 \degree$ at the trigger time, decreases down to a minimum value
of $\simeq 35 \degree$ at $t_0 + 60$ s, and then increases again
up to $\simeq 70 \degree$ at $t_0 + 140$ s. Assuming an FoV of $60
\degree$ radius for the GRID, GRB 100724B is contained within the
FoV between $t_0 + 6$ s and $t_0 + 125$ s (thus the measured
duration is 119 s) and then again between $t_0 + 410 $ s and $t_0
+ 529 $ s.

We selected the GRID data of GRB 100724B in the time interval
defined above (from $t_0 + 6$ s until $t_0 + 125$ s) and in a $12
\degree$ radius from the source position. To exclude the
contamination from the Earth albedo, we selected photons with an
angular distance greater than $90 \degree$ from the Earth centre.
For GRB 100724B we classified the events in the photon list with
the \texttt{F4} filter \citep[][and references
therein]{Giuliani_et_al_2008}, which provides the highest
statistic. Using these cuts and selections we extracted 57 events
from GRB 100724B with reconstructed energy from 22 MeV to 3.5 GeV
(see fig. \ref{fig:MCAL-GRID_lightcurve}). The GRB is localised by
GRID at $RA=124.3 \degree$, $dec=78.5 \degree$ with a 95 \%
confidence level uncertainty of $1.5 \degree$ radius (including a
systematic contribution of $0.1 \degree$), at $3.0 \degree$ from
the centre of the IPN error region \citep[reported
by][]{Guiriec_et_al_2011}.

\subsection{Lightcurves and cross-correlation}

\begin{figure}[h!]
\centering
\includegraphics[width=9. cm]{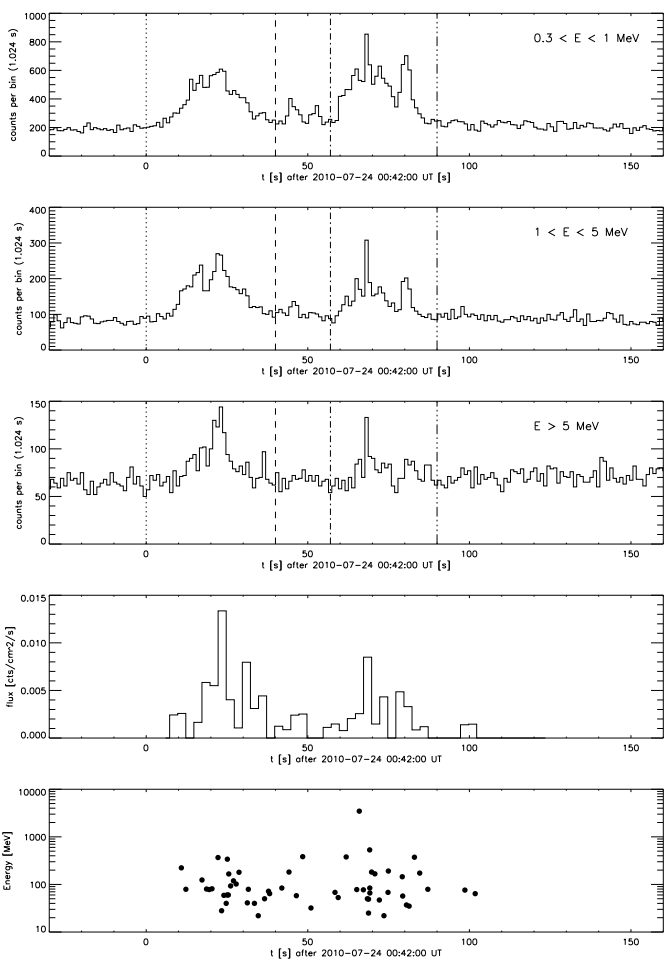}
\hfill \caption{Lightcurves of GRB 100724B: from the top down MCAL
0.3 -- 1 MeV, MCAL 1 -- 5 MeV, MCAL above 5 MeV, GRID flux
corrected for the effective area and GRID scatter plot of energy
vs time.} \label{fig:MCAL-GRID_lightcurve}
\end{figure}

The statistics of the MCAL data allowed us to extract the
lightcurves in three energy bands: 0.3 -- 1 MeV, 1 -- 5 MeV, and
above 5 MeV. We accumulated the GRID lightcurve by converting the
detected counts into flux using the instrument's effective area as
a function of the off-axis angle, depending on time. The
superposition of the MCAL and GRID lightcurves of GRB 100724B is
shown in fig. \ref{fig:MCAL-GRID_lightcurve}.

GRB 100724B is significantly detected by MCAL up to $\simeq 80$
MeV energy. We searched backward for possible precursors in the
MCAL available data, until $t_0 - 40$ s, but we did not find any
at a significance level of $5 \sigma$.  The time intervals
$T_{90}$ and $T_{50}$ enclose the burst counts from 5 \% to 95 \%,
and 25 \% to 75 \% of the total, respectively, and are
calculated using the standard procedure outlined by
\citet{Koshut_et_al_1996}. From the counts over the full MCAL
energy range (350 keV -- 100 MeV), $T_{50} = 50.6 \pm 1.1$ s and
$T_{90} = 96.4 \pm 10.8$ s. Our $T_{90}$ is shorter, but still
compatible, with the 111.6 s value reported by GBM
\citep{GCN_10977} for the 50 -- 300 keV energy band. It is
important to point out here that the MCAL response, hence the
measure of the burst duration in the MeV band, is negligibly
affected by the \textit{AGILE} spinning mode.

We studied the shape of the MCAL lightcurve by selecting the data
in the three energy ranges listed above (0.3 -- 1 MeV, 1 -- 5 MeV,
and above 5 MeV). The lightcurve between $t_0$ and $t_0 + 96.4$ s
is composed of two bumps, both detected in the whole energy range
and characterised by internal structure, interleaved by an
interbump pause region, where two softer and fainter peaks are
present, the first one only visible below 5 MeV and the second one
only below 1~MeV (see fig. \ref{fig:MCAL-GRID_lightcurve}). The
time series in the three energy bands are qualitatively similar,
with the two main bumps peaking at similar values of time. Above 5
MeV most of the internal structure in the bumps is not
statistically significant above the background, for example the
narrow peak at the end of the second bump (at $\simeq t_0 + 80$
s), while a narrow peak is present only above 5 MeV around $\simeq
t_0 + 35$ s but with a significance of only $3 \sigma$.

\begin{figure}[h!]
\centering
\includegraphics[width=5.5 cm, angle=90]{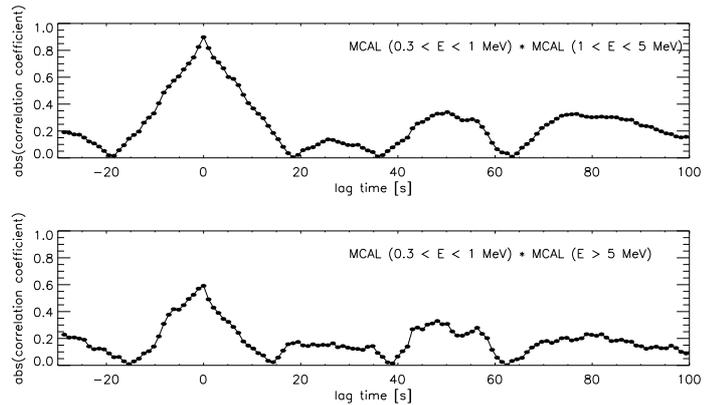}
\hfill \caption{Cross-correlation of GRB 100724B between the MCAL
lightcurve at 0.3 -- 1 MeV and the lightcurves from 1 to 5 MeV
(top panel) and above 5 MeV (bottom panel).}
\label{fig:MCAL_cross-cor}
\end{figure}

We studied the presence of energy-dependent lags in GRB 100724B by
computing the cross-correlation function (CCF) of the time series
in different energy ranges. The CCF is defined in most textbooks
of statistical analysis \citep[see for
example][]{Bevington_Robinson_2003}. In fig.
\ref{fig:MCAL_cross-cor} we show the absolute value of the CCF
between the MCAL lightcurve in the 0.3 - 1 MeV energy band and the
lightcurves in 1 -- 5 MeV (top panel) and above 5 MeV (bottom
panel). We use a time lag between $-30$ s and $+160$ s and a bin
size of 1.024 s (the same as in fig.
\ref{fig:MCAL-GRID_lightcurve}), the shortest allowed by the
statistics of the data. The peak in the CCF is taken as the
measure of spectral lag \citep{Norris_2002}. We can see from fig.
\ref{fig:MCAL_cross-cor} that the time lag in the MCAL lightcurves
is less than 1.024 s.

The position of the maximum emission in the first bump does not
change significantly in the three energy bands, as demonstrated
with the cross-correlation function. Conversely, the first bump
width shows variations as a function of energy, measured by
assuming a symmetric Gaussian shape. From the Gaussian fit we
found, again, that the position does not change significantly with
energy, while the width, consistent in the first and second bands,
is smaller in the highest energy interval. The Gaussian function
is just a qualitative model for the bump shape, and the high
values of the reduced chi square, which are not formally
acceptable, are dominated by the internal peak structure.
Similarly, the position of the maximum emission in the second main
bump does not change significantly in the three energy bands but,
in this case, the bump shape is dominated by the superposition of
many overlapping peaks and thus cannot be modelled using the same
method as applied to the first bump.

\begin{figure}[h!]
\centering
\includegraphics[width=8. cm, angle=90]{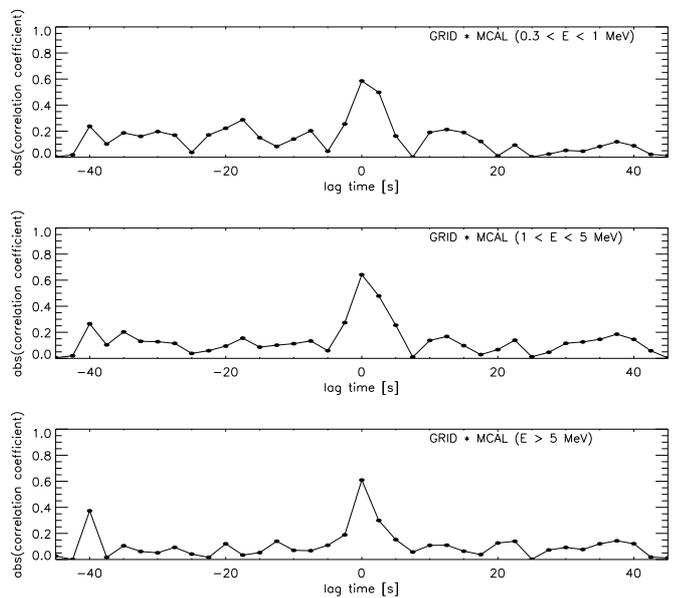}
\hfill \caption{Cross-correlation of GRB 100724B between the
lightcurves of the GRID and of the MCAL between 0.3 and 1 MeV (top
panel), between 1 and 5 MeV (middle panel) and above 5 MeV (bottom
panel).} \label{fig:MCAL-GRID_cross-cor}
\end{figure}

The GRID lightcurve shows two broad bumps, at the same position as
in MCAL, without evident features in the interbump region (see
fig. \ref{fig:MCAL-GRID_lightcurve}). The average background level
in the GRID time series around the time interval of GRB 100724B is
$0.12 \; \mathrm{cts \; s^{-1}}$, corresponding to a flux of $4
\times 10^{-4} \; \mathrm{ph \; cm^{-2} \; s^{-1}}$ when corrected
for the effective area at 100 MeV for an average off-axis angle of
$50 \degree$.

The onset of the gamma ray emission does not show any significant
delays with respect to the hard X-ray band; in fact the first
photon is detected by GRID at $t_0 + 10.9$ s, during the rise of
the first bump in MCAL, and has an energy of $\simeq 220$ MeV. The
highest energy photon ($\simeq 3.5$ GeV) is detected by GRID
immediately before the maximum of the second bump in the MCAL
lightcurve. The narrow peak around $\simeq t_0 + 35$ s in the MCAL
lightcurve above 5 MeV is also evident in the GRID data.

The absolute value of the CCF between the time series of the GRID
and MCAL (without background subtraction) in the three energy
bands (0.3 - 1 MeV, 1 -- 5 MeV, and above 5 MeV) is shown in fig.
\ref{fig:MCAL-GRID_cross-cor}. In the cross-correlation we used
for all the lightcurves the same time bin of 2.5 s, which is the
shortest value allowed by the statistics of the GRID data and the
same one as used for the lightcurve in fig.
\ref{fig:MCAL-GRID_lightcurve}. Since the GRB is within the GRID
FoV only from $t_0 + 6$ s until $t_0 + 125$ s, we used a time lag
range between $-45$ s and $+45$ s, which is different from the CCF
at lower energy (in fig. \ref{fig:MCAL_cross-cor}), in which data
from a wider time interval can be used. Again, the correlation
coefficient is maximum at a zero time lag, indicating the absence
of a spectral lag longer than 2.5~s between the time series of
GRID and MCAL. This result agrees with the comparison between the
MCAL and GRID lightcurves, indicating that the position of the
maximum emission is coincident in the first bump at $t_0 + 22.5$
s, as well as in the second one at $t_0 + 67.5$ s (see fig.
\ref{fig:MCAL-GRID_lightcurve}). The width of the first bump in
GRID is similar to MCAL ($\simeq 10$~s at half maximum), while the
width of the second one can only be estimated with difficulty
because of the low-quality statistics.

Recent observations of GeV-bright GRBs have shown that gamma rays
are detected in an extended emission prompt phase, lasting well
after $T_{90}$ and up to tens or even hundreds of seconds after
trigger. The prototype of this behaviour is represented by GRB
940217 \citep{Hurley_et_al_1994}, and recent cases are, for
example, GRB 080514B \citep{Giuliani_et_al_2008} detected by
\textit{AGILE}, GRB 090510
\citep{Giuliani_et_al_2010,Ackermann_et_al_090510,Kumar_Barniol-Duran_2010}
studied by \textit{AGILE} and \textit{Fermi}, and GRB 080825C
\citep{Abdo_et_al_080825C} observed by \textit{Fermi}. The study
of this phenomenology in GRB 100724B is complicated by
\textit{AGILE} operating in a spinning mode, consequently the
burst ``exits'' from the GRID FoV on $t_0 + 125$~s, and the
subsequent ``transit'' lasts between $t_0 + 410 $ s and $t_0 + 529
$ s. We extracted the GRID data in this second time interval,
using the same cuts and selections of the first interval
(represented in fig. \ref{fig:MCAL-GRID_lightcurve}), and we found
only 11 events. When taking the same background level measured
before trigger into account, thanks to the stability of the GRID
particle background in the equatorial low Earth orbit
\citep{Tavani_et_al_2009}, the signal-to-noise ratio in this case
is just $1.0 \sigma$. The significance level remains similar if we
reduce the extraction radius down to $10 \degree$, obtaining $1.4
\sigma$, or even $8 \degree$, resulting in $1.5 \sigma$.

\subsection{Time-resolved spectral analysis}

The time-integrated spectrum of MCAL between $t_0$ and $t_0+90$ s
can be fit in the energy range 0.5 -- 80 MeV with a single
powerlaw and the resulting photon index is $-2.13^{+0.05}_{-0.04}$
with reduced $\chi^2$ = 0.92 and 61 degrees of freedom (dof). The
estimated fluence in the same energy and time intervals is $(2.3
\pm 0.2) \times 10^{-4} \; \mathrm{erg \; cm^{-2}}$. All reported
errors are at the 90~\% confidence level.

\begin{figure}[h!]
\centering
\includegraphics[angle=-90, width=10. cm]{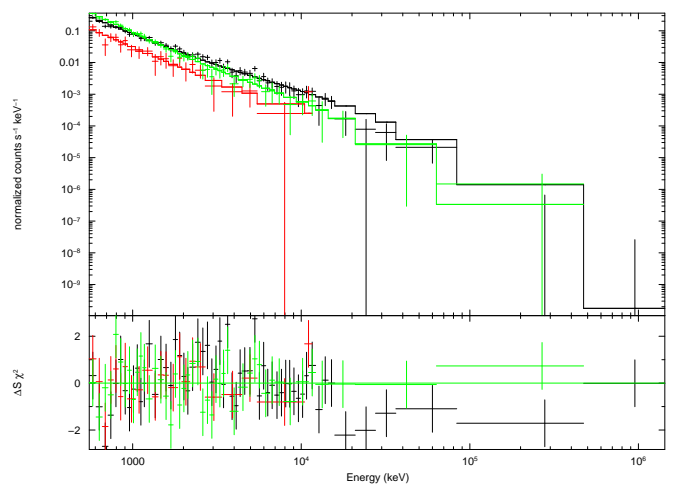}
\hfill \caption{Superposition of the MCAL spectra in the three
time intervals defined in Table \ref{table:MCAL_spectra}: $t_0$ --
$t_0 + 40$ s (interval A, black in colour version), $t_0 + 40$ s
-- $t_0 + 57$ s (interval B, red in colour version) and $t_0 + 57$
s -- $t_0 + 90$ s (interval C, green in colour version).}
\label{fig:MCAL_spectra}
\end{figure}

We identified three time intervals to study the time-resolved
spectrum of the GRB prompt emission, listed in Table
\ref{table:MCAL_spectra}: the first main bump (from $t_0$ until
$t_0 + 40$ s, hereafter interval A), the interbump region (from
$t_0 + 40$ s until $t_0 + 57$ s, hereafter interval B), and the
second main bump (from $t_0 + 57$ s until $t_0 + 90$ s, hereafter
interval C). We fitted the MCAL spectra of the three intervals in
the energy range between 0.5 and 100 MeV with a simple powerlaw
and the results are reported in Table \ref{table:MCAL_spectra}. We
find evident spectral evolution at a significance level of $4.0
\sigma$, with the first bump harder than the second one (see fig.
\ref{fig:MCAL_spectra}), as qualitatively expected from the
comparison of the bump amplitude and integrated flux in the
lightcurves at different energies (in fig.
\ref{fig:MCAL-GRID_lightcurve}). The first bump is also harder
than the interbump region, but in this case a significance level
lower than $1 \sigma$ is found, owing to the lower statistics in
the spectrum.

Fitting the time integrated spectrum of the GRB data only in GRID
with a powerlaw we obtain a photon index of
$-2.04^{+0.31}_{-0.14}$, with a total fluence between 22 MeV and
3.5 GeV of $0.25 \pm 0.05 \; \mathrm{ph \; cm^{-2}}$,
corresponding to $(4.7 \pm 0.9) \times 10^{-5} \; \mathrm{erg \;
cm^{-2}}$. We also fitted the joint spectrum of MCAL and GRID with
the least squares method using a program specifically developed
for the GRID \citep[see for example][]{Pucella_et_al_2008}. With
the available statistics we could only arrange the GRID data in
three energy bands (22 -- 70 MeV, 70 -- 200 MeV, and 200 -- 3500
MeV). The joint fit confirms that a single powerlaw, with the same
photon index of $-2.13$ measured in the MCAL time-integrated
spectrum, is an adequate model for the spectrum from 500 keV up to
3.5 GeV.

GRB 100724B turns out to be the brightest GRB detected by
\textit{AGILE} in gamma rays during its operations in space. The
statistics of the GRID data are not good enough to appreciate a
possible spectral evolution in the GeV energy band. We verified
the possibility of detecting the same spectral evolution measured
by MCAL with the GRID by estimating the significance of a
variation in the hardness ratio. Whenever we imposed the same
values of the photon indices as measured by MCAL, $-2.01$ in the
first time interval and $-2.35$ in the third one, and we compared
the variation in the resulting hardness ratio with the uncertainty
from our data, we find that the variation is at a significance
level of $1.3 \sigma$, thus we cannot draw any serious conclusion.

\subsection{Minimum Lorentz factor}

The mere fact that the spectrum of a GRB does not show any cutoff
in the GeV energy interval can be used to estimate the minimum
Lorentz factor $\Gamma_{min}$ of the prompt emission fireball
because the optical depth of the interactions of gamma rays on
low-energy photons is small
\citep{Baring_Harding_1997,Lithwick_Sari_2001}.

\begin{figure}[h!]
\centering
\includegraphics[width=7. cm, angle=90]{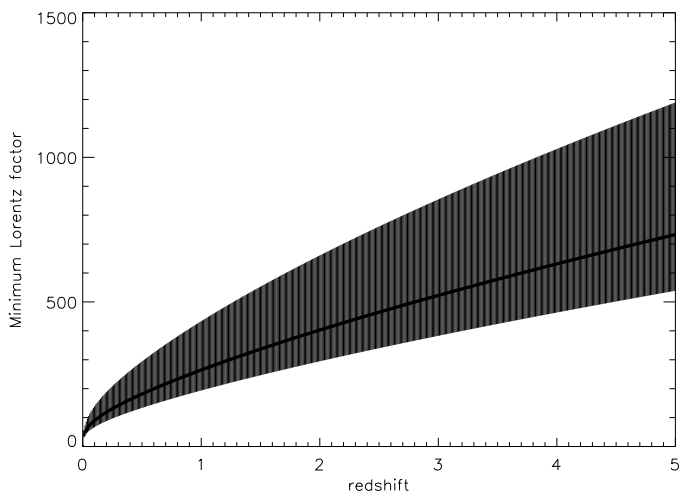}
\hfill \caption{Estimation of the minimum Lorentz factor, with
uncertainty, following the method in the Supporting Online
Material of the paper by \citet{Abdo_et_al_080916C}, from the
highest energy photon detected by GRID ($3.5 \pm 1.7$ GeV), the
powerlaw photon index ($-2.04^{+0.31}_{-0.14}$) and the typical
variability timescale of the prompt emission from the MCAL
lightcurve ($1.2 \pm 0.1$ s).} \label{fig:Lorentz_min}
\end{figure}

We estimated the minimum value of the bulk Lorentz factor for GRB
100724B following the formulae reported in the supporting online
material of the paper by \citet{Abdo_et_al_080916C}. The GRID
detected a photon of energy as high as $3.5 \pm 1.7$ GeV from GRB
100724B on $t_0 + 65.9$ s. In the estimation of the minimum
Lorentz factor, \citet{Ackermann_et_al_090510} assume the full
width at half maximum (FWHM) of the shortest time pulse as the
variability timescale of the emission. Near the arrival time of
the highest energy photon ($t_0 + 65.9$ s), the MCAL lightcurve
shows a peak with FWHM of $1.2 \pm 0.1$ s, which we assume as the
timescale of the burst variability. Since GRB 100724B lacks a
redshift, we estimate $\Gamma_{min}$ as a function of redshift, in
an interval between 0 and 5. Assuming a standard cosmology with
$H_0 = 70 \; \mathrm{km \; s^{-1} \; Mpc^{-1}}$, $\Omega_M= 0.3$,
and $\Omega_{\Lambda}= 0.7$, we obtain a minimum Lorentz factor
ranging between $\simeq 50$ and $\simeq 700$ depending on
redshift, as shown in fig. \ref{fig:Lorentz_min}. The uncertainty
on $\Gamma_{min}$ is dominated by the uncertainty on the energy
reconstruction by GRID, $\Delta E / E \sim 1$
\citep{Tavani_et_al_2008}. Other ingredients of the overall
uncertainty are the errors on the photon index and variability
time scale.

\section{Discussion and conclusions}
\label{sec:Discussion}

\subsection{Fluence}

GRB 100724B is characterised by a high fluence, as measured by
\textit{AGILE} both in hard-X rays, ($2.3 \pm 0.2) \times 10^{-4}
\; \mathrm{erg \; cm^{-2}}$ in 0.5 -- 80 MeV, and in gamma rays,
$(4.7 \pm 0.9) \times 10^{-5} \; \mathrm{erg \; cm^{-2}}$ above 22
MeV. We note here that we observed only the initial $\simeq 125$ s
of the GRB gamma ray emission, thus we can provide only a lower
limit to the fluence in the GRID energy band. Following the
analysis by \citet{Guiriec_et_al_2011} of the GBM observation of
GRB 100724B, the burst fluence between 8 keV and 10 MeV is $4.2
\times 10^{-4} \; \mathrm{erg \; cm^{-2}}$, roughly corresponding
to $\simeq 2.7 \sigma$ ($\simeq 99.3 \; \%$) in the distribution
of the fluence values measured by GBM for 121 GRBs (on October
2009) and considered in the analysis by
\citet{Ghisellini_et_al_2010}. Rescaling the gamma ray spectrum
measured by the AGILE/GRID to the same energy band (0.1 -- 100
GeV) of \citet{Ghisellini_et_al_2010}, we obtain an average energy
flux of $5.1 \times 10^{-7} \; \mathrm{erg \; cm^{-2} \; s^{-1}}$,
fully consistent with the results of that paper and indicating
that GRB 100724B belongs to the same population.

\subsection{Lightcurves and cross-correlation}

While the measures of the burst duration and $T_{90}$ in the hard
X-ray band are negligibly affected by the \textit{AGILE} operating
in a spinning mode, thanks to the almost all-sky FoV of the MCAL,
we can just put a lower limit at 119 s on the duration of the GeV
emission since the GRB went out of the GRID FoV at $\simeq t_0 +
125$ s and its emission was fainter than the instrument
sensitivity when the burst was back inside, on $t_0 + 410$ s (i.
e. 285 s later). For this reason we cannot draw any serious
conclusion about the duration of the GRB in gamma rays from the
\textit{AGILE} data.

The cross-correlation of the lightcurves of GRB 100724B without
background subtraction and in various energy ranges (0.3 -- 1 MeV,
1 -- 5 MeV, 5 -- 80 MeV, and 22 MeV -- 3.5 GeV) shows two peculiar
characteristics: the simultaneous onset of the gamma ray and hard
X-ray emissions and the coincidence of the peak position in the
time series. This is demonstrated by the absence of spectral lag
between the MCAL lightcurves at different energies, with a bin
size of 1.024 s (see fig. \ref{fig:MCAL_cross-cor}) and between
the GRID lightcurve and the MCAL ones in three energy ranges, with
a bin size of 2.5 s, mainly limited by the statistics of the GRID
data (see fig. \ref{fig:MCAL-GRID_cross-cor}). Moreover, the first
photon is detected by the AGILE/GRID at 10.9 s after trigger,
during the rise of the burst emission in the MeV band. The MCAL
and GRID time series remain simultaneous also if we accumulate the
data from the whole energy intervals of each instrument, showing
that the position of the main bumps is the same, without
significant time lags.

The absence of time lag makes GRB 100724B unusual among the events
observed so far in gamma rays. In fact, some similarities in the
lightcurve can be found in GRB 090217A
\citep{Ackermann_et_al_090217A}, but in that case the first
$\simeq 3$ s in the gamma ray lightcurve are considerably fainter
than in hard X-rays, and in GRB 080916C
\citep{Abdo_et_al_080916C}, but its gamma ray emission starts at
the second peak, $\simeq 3.6$ s after trigger. The lack of gamma
rays during the first peak at MeV of GRB 080916C and the presence
of them in the second one and after is used as an argument by
\citet{Abdo_et_al_080916C} to suggest that the two peaks may
origin from spatially distinct regions or from two pairs of
colliding shells that are different in physical conditions and
hardness. Following the same reasons, we can argue that the GeV
and MeV emissions of GRB 100724B may originate in a common
spatial region or, in the internal-shock scenario, from the
collision of the same pair of shells. Similarly,
\citet{Maxham_Zhang_Zhang_2011} consider the simultaneity of the
GeV and MeV emission and the single spectral shape across the wide
energy band as a more definite argument in favour of the internal
shock origin of the gamma ray component, assumed as a spectral
extension to lower energies of the MeV emission.

\subsection{Time resolved spectrum}

GRB 100724B is characterised by an evident spectral evolution in
the MeV energy band. In fact, a significant hard-to-soft variation
in the the spectrum from the first ($t_0$ -- $t_0 + 40$ s) to the
second bump ($t_0 + 57$ s -- $t_0 + 90$ s) is detected by the
AGILE/MCAL (see fig. \ref{fig:MCAL_spectra} and Table
\ref{table:MCAL_spectra}), while the statistics in the gamma ray
band does not allow appreciation of a similar evolution. The burst
spectral evolution appears stronger if we look at the Konus-Wind
results\footnote{\texttt{http://www.ioffe.ru/LEA/GRBs/GRB100724\_T02526/}}
\citep{GCN_10981} in the 18 -- 1160 keV band, in which the second
broad bump ($t_0 + 50$ s -- $t_0 + 85$ s) is brighter than the
first one ($t_0$ -- $t_0 + 25$ s) and there is also a soft bump,
extending approximately from $t_0 + 100$ s until $t_0 + 140$ s and
visible only below 70 keV energy, therefore not detected by MCAL.
The same feature is also detected by GBM below 200 keV
\citep[][]{Guiriec_et_al_2011}.

The photon index of $-2.13^{+0.05}_{-0.04}$ obtained from the fit
of the time-integrated MCAL spectrum (see Table
\ref{table:MCAL_spectra}) is in good agreement with the highest
energy photon index from the Band function measured at
$-2.00^{+0.07}_{-0.09}$ by Konus-Wind in the 20 keV -- 10 MeV
energy range \citep{GCN_10981}, although on a longer time interval
(from $t_0 + 6$ s to $t_0 + 235$ s), and with the value of $-1.99
\pm 0.01$ measured by \citet{Guiriec_et_al_2011} from the
Fermi/GBM data. For this burst the peak energy is measured at
$369_{-37}^{+42}$ keV by Konus-Wind \citep{GCN_10981} and at $352
\pm 6$ keV by Fermi/GBM \citet{Guiriec_et_al_2011}.

In this analysis of the Fermi/GBM data, \citet{Guiriec_et_al_2011}
find that the fit of the spectra in the 8 keV -- 40~MeV energy
band significantly improves by including a thermal component of
$kT = 38.14 \pm 0.87$ keV. In this case, the high-energy photon
index $\beta$ of the Band function is softer, $-2.11 \pm 0.02$.
With an energy threshold of $ \sim 300$ keV, the additional
spectral component found by Fermi/GBM at $kT = 38.14 \pm 0.87$ keV
cannot be detected by the AGILE/MCAL. Moreover, the reduced chi
square value of 0.92 from the MCAL integrated spectrum implies
that a simple powerlaw is an adequate model and does not allow us
to add other components. Nevertheless, we tried to introduce a
thermal component in the spectrum, but it does not affect the flux
above 500 keV and does not improve the reduced chi square (that in
fact increases from 0.92 to 0.95). For these reasons we can
conclude that the MCAL spectum is adequately fit by a simple
powerlaw.

The time-resolved spectral analysis by \citet{Guiriec_et_al_2011}
also shows substantial variations in the $E_{peak}$ parameter of
the Band function from $\sim 90$ keV up to $\sim 1100$~keV. In the
first half of our interval A, $E_{peak}$ is higher ($\sim 1100$
keV) and then decreases down to $\sim 500$ keV in the B and C time
intervals. In the AGILE/MCAL energy range, the spectrum can only
be fitted with a powerlaw, thus the $E_{peak}$ parameter cannot be
directly measured. However, the MCAL spectrum is harder (with a
lower absolute value of the powerlaw photon index) in the A
interval and becomes softer later on, when the absolute value of
the photon index increases. The evolution of the photon index in
MCAL corresponds to the high-to-low time variation in the
$E_{peak}$ parameter measured by GBM.

Similar to GRB 090902B \citep{Abdo_et_al_090902B}, the gamma ray
emission of GRB 100724B shows large amplitude fluctuations in the
GeV component (of more than a factor of five, see fig.
\ref{fig:MCAL-GRID_lightcurve}) that can be explained with
difficulty in the early afterglow scenario and which are generally
attributed to the prompt emission \citep{Abdo_et_al_090902B}. As
in the case of GRB 080916C \citep{Abdo_et_al_080916C}, the
adequacy of a single spectral model from hundreds of keV up to few
GeV energies is compatible with a non-thermal synchrotron origin
of the radiation. The high-energy electrons should, however,
upscatter via inverse Compton effect the photons emitted by the
synchrotron up to GeV energies, thus producing a synchrotron
self-Compton (SSC) component that should be detected by the
AGILE/GRID. The lack of this gamma ray component in the spectrum
of GRB 100724B can be explained, as for 080916C, if the magnetic
energy density is much higher than the electron density, thus the
SSC is suppressed, or if the SSC peaks far above tens of GeV, so
cannot be detected by the AGILE/GRID \citep[see the supporting
online material of the paper by][for further
information]{Abdo_et_al_080916C}.

From the cross-correlation analysis and the time-resolved spectral
analysis, we find that GRB 100724B is characterised by the
presence of a spectral evolution and absence of a spectral lag.
These features are rather unusual and are, for example, not found
in the other GRBs detected so far in gamma rays by \textit{AGILE}.

\subsection{Minimum Lorentz factor}

Estimated on the hypothesis of negligible optical thickness at GeV
energy, the minimum Lorentz factor of the GRBs detected in gamma
rays ranges between $\simeq 600$ and $\simeq 900$ for the long GRB
080916C \citep{Abdo_et_al_080916C} and from $\simeq 700$ to
$\simeq 1300$ for the short GRB 090510
\citep{Ackermann_et_al_090510}. In the redshift interval from 0 to
5, GRB 100724B has a slightly lower minimum Lorenzt factor,
between $\simeq 50$ and $\simeq 700$ (see fig.
\ref{fig:Lorentz_min}). While the lowest value is similar to the
one calculated for GRB 080916C and GRB 090510 for $z=0$, we find
only $\simeq 700$ at $z=5$, depending on whether the variability
timescale in \textit{AGILE} is similar to the one measured by
\textit{Fermi}, but the maximum detected photon energy is less due
to the lower efficiency of \textit{AGILE} above $\simeq 1$ GeV.

\begin{acknowledgements}

\textit{AGILE} is a mission of the Italian Space Agency, with
co-participation of INAF (Istituto Nazionale di Astrofisica) and
INFN (Istituto Nazionale di Fisica Nucleare). This work was
partially supported by ASI grants I/R/045/04, I/089/06/0,
I/011/07/0 and by the Italian Ministry of University and Research
(PRIN 2005025417). INAF personnel at ASDC are under ASI contract
I/024/05/1. This research has made use of NASA's Astrophysics Data
System. Finally, we acknowledge the contribution of the anonymous
referee, who stimulated us to substantially improve the quality of
the paper.

\end{acknowledgements}

\bibliographystyle{aa} 
\bibliography{GRB100724B_2010}

\begin{table*}[p!]
\caption{Fit results of the MCAL spectra of GRB 100724B, with times relative to the trigger time $t_0$ corresponding to 2010-07-24 00:42:00 UT.}             
\label{table:MCAL_spectra}      
\centering                          
\begin{tabular}{c c c c c c}        
Time interval & photon index & $\chi_r^2$ (dof) & flux 0.5 -- 100 MeV\\    
              &              &                  & [$\mathrm{erg \; cm^{-2} \; s^{-1}}$] \\
\hline                        
whole ($t_0$, $t_0 + 90$ s)    & $2.13^{+0.05}_{-0.04}$ & 0.923 (61) & $2.38 \times 10^{-6}$ \\ 
A ($t_0$, $t_0 + 40$ s)        & $2.01 \pm 0.04$        & 1.226 (54) & $3.35 \times 10^{-6}$ \\ 
B ($t_0 + 40$ s, $t_0 + 57$ s) & $2.19^{+0.26}_{-0.19}$ & 0.661 (21) & $9.18 \times 10^{-7}$ \\ 
C ($t_0 + 57$ s, $t_0 + 90$ s) & $2.35^{+0.08}_{-0.07}$ & 0.663 (42) & $2.24 \times 10^{-6}$ \\ 

\hline                                   
\end{tabular}
\end{table*}

\end{document}